\documentclass[a4paper,11pt]{article}
\usepackage{jheppub}
\usepackage[T1]{fontenc}

\title{Nonperturbative QED vacuum birefringence}
\author{V.I.Denisov}
\author{E.E.Dolgaya}
\author{and V.A.Sokolov}
\affiliation{Physics Department, Moscow State University, Moscow,
119991, Russia}
\emailAdd{vid.msu@yandex.edu}
\emailAdd{eedolgy@gmail.com}
\emailAdd{sokolov.sev@inbox.ru}
\abstract{In this paper we represent nonperturbative
calculation for one-loop Quantum Electrodynamics (QED) vacuum birefringence
in presence of  strong magnetic field.
The dispersion relations for electromagnetic wave
propagating in strong magnetic field point to retention of
vacuum birefringence even in case when the field strength
greatly exceeds Sauter-Schwinger limit. This gives a possibility to extend some
predictions of perturbative QED such as electromagnetic waves delay in pulsars neighbourhood
or wave polarization state changing (tested in  PVLAS)
to arbitrary magnetic field values. Such expansion
is especially important in astrophysics because magnetic fields
of some pulsars and magnetars  greatly exceed quantum magnetic field
limit, so the estimates of perturbative QED effects
in this case require clarification.}


\begin{document}
\maketitle
\flushbottom

\section{Introduction}\label{sec:intro}
From the very beginning,  radiative corrections
to the QED Lagrangian coming
from the vacuum fluctuations
have been the subject of a great interest.
Such corrections, accounted on background of constant and homogeneous
electromagnetic field, lead to Heisenberg-Euler effective Lagrangian~\cite{a1}.
The scale parameter for nonlinearities in  Heisenberg-Euler electrodynamics
(characteristic quantum electrodynamic induction or Schwinger limit for electric field)
$B_c=E_c=m^2c^3/e\hbar=4.41 \cdot 10^{13}$G distinguishes different regimes of the theory.

The perturbative or post-Maxwellian regime takes place
for most of the available in contemporary laboratory electromagnetic
fields $E, B\ll E_c, B_c$.
In this case the Heisenberg-Euler Lagrangian can be
expanded in power series of electromagnetic field tensor
invariants. The perturbative regime of QED is well understood
and many of its predictions  have been experimentally observed.
For instance, electron anomalous magnetic moment
remains a great example of unprecedented correspondence between
theoretical and experimental results~\cite{a2,a3}.
Some other predictions such as the Delbr\"{u}ck scattering~\cite{a4},
Lamb shift~\cite{a5} and photon splitting~\cite{a6,a7}
are also well-established.
Another manifestation of perturbative  QED lies in the phenomenon of
vacuum  behavior like a polarizable medium with
the cubic nonlinearity in constitutive relations.
Such semiclassical description leads to birefringence of electromagnetic waves
in vacuum in presence of external field.
The attempt of vacuum birefringence experimental observation
was made in PVLAS experiment~\cite{a8,a9}. The measurement of refractive
indexes induced by vacuum birefringence was based on
detecting rotation of electromagnetic wave polarization
when it propagated in dipole magnetic field
with induction $B_{ext} = 2.5 \cdot 10^4$G.
The results of PVLAS set  a  new limit on vacuum
magnetic birefringence above the level pointed out by QED~\cite{a10,a9}.
According to the authors, the discrepancy can be explained
beyond the Standard Model by interaction with axiones.
The experiment also bounded the coupling constant of axion-like particles and
photons~\cite{a10,a9}.
Since the vacuum birefringence
is a very small macroscopic
quantum effect  it's
detection  needs  strong enough magnetic field source,
which is difficult to obtain in laboratory.
At the same time, usage of compact astrophysical objects as
natural magnetic fields sources provides wide
opportunities for vacuum birefringence investigation
due to the fact that for many pulsars and manetars the
magnetic field is close to or even exceeds $B_c$. Vacuum
birefringence in this case can be detected
by measuring X- and gamma- ray polarization
passing the region of the strong magnetic
field near the pulsar. Due to the difference in
wave propagation velocities induced by vacuum
birefringence the time lag between the arrival
of the fast and the slow mode to the detector
is proportional to vacuum
refractive index difference for each
polarization mode. The calculations in perturbative QED regime
show the detectability of the effect~\cite{a11} and give the
value for the time lag $\Delta t\sim 10^{-7}$s.
The results of calculations are valid when
the pulsar field $B<B_c$, however the existence of pulsars
with the overcritical field (for instance B1509-58 with the $B\sim 1.5\cdot 10^{14}$G)
provides an attractive possibility to enhance the estimates for
the time lag. Such calculations require of vacuum birefringence consideration outside the perturbative  regime.

Nonperturbative regime  of QED  arises when we consider
external fields which values are
close to or exceed the scale parameter of the
Heisenberg-Euler electrodynamics
$E, B \sim E_c, B_c$.
This regime shows one of the most
amazing properties -- vacuum instability due
to electron-positron pair production. The
effect is expected~\cite{a14} when the electric field
exceeds so-called Schwinger limit $E>E_c$
and, although it has never been observed directly, the
advances of high-intensity laser physics and
implementation of such projects as ELI, XFEL
and others~\cite{a12,a13}
give a great promise in  this area.
In general, QED in nonperturbative regime
is poorly investigated and
provides  new challenges both
in theoretical and experimental research.

In this paper we will focus on one-loop QED
vacuum birefringence in nonperturbative regime.
In addition to entirely quantum technic
proposed in~\cite{a15,a16}, we represent
a semiclassical approach. The maim aim of
our research is to expand  vacuum
birefringence predictions for  different experiments
to values of strong magnetic fields close to
character quantum electrodynamic induction.
The paper is organized as follows: in Sec.\ref{Sec2} we
derive constitutive relations for QED, Sec.\ref{Sec3} is devoted to the
weak electromagnetic wave propagation in strong magnetic field,
in Sec.\ref{Sec4} we discuss QED birefringence expansion following
from obtained dispersion relations and possibilities for it's observation and in conclusive Sec.\ref{Sec5} we summarize our results.

\section{Constitutive relations for nonperturbative QED}\label{Sec2}
The Lagrangian in QED is represented as a series of
corrections to Maxwell electrodynamics.
One-loop QED considers only the first non-vanishing correction
which in non-perturbative regime has the following form~\cite{b1}:
\begin{equation}\label{Lagrange_main}
L_1=-{\alpha B_c^2\over 8\pi^2}\int\limits_0^\infty {e^{-s}\over s^3}
\Big[sa \ \cot(sa)\cdot
sb \ \coth(sb) -1 -{s^2\over 3}(b^2 - a^2)\Big]ds,
\end{equation}
where $\alpha=e^2/\hbar c$ -- is a fine structure constant and the parameters $a$ and $b$  are expressed by the
electromagnetic field components:
\begin{equation}\label{ab}
a=-{i\over \sqrt{2} B_c}\Big(\sqrt{F+iG}-\sqrt{F-iG}\Big),
\quad b={1 \over \sqrt{2} B_c}\Big(\sqrt{F+iG}+\sqrt{F-iG}\Big),
\end{equation}
where the notations
$F=({\bf B}^2-{\bf E}^2)/2$ and $G=({\bf E} {\bf B})$ were used for brevity.
Strictly speaking, the Heisenberg-Euler Lagrange function
is valid  only in  case of constant and homogeneous
background fields, at least at the typical
scale of Compton wavelength $\lambda_c = h/mc$.
Also it should be noted that the auxiliary parameters $a$ and $b$ of
Heisenberg-Euler Lagrangian (\ref{Lagrange_main}) are special
because  $\pm a$ and $\pm b$ are the eigenvalues of the
electromagnetic field tensor $F_{ik}$  when field value is constant
and this makes the problem of QED radiative
corrections exactly solvable \cite{arx1}.

In semiclassical approach  Heisenberg-Euler theory
can be interpreted as nonlinear
electrodynamics of continuous media with
special constitutive relations, in which
polarization ${\bf P}$ and magnetization ${\bf M}$ induced
by the external fields ${\bf E}$ and ${\bf B}$,
can be expressed from the Lagrangian:
\begin{equation}\label{PM_general}
{\bf P}={\partial L_1\over \partial {\bf E}},
\qquad {\bf M}={\partial L_1\over \partial {\bf B}}.
\end{equation}
In order to calculate the explicit values for this vectors, it is
useful to introduce some auxiliary relations:
\begin{equation}\label{Aux_rel}
{\partial a \over \partial {\bf E}}={\partial b \over \partial {\bf B}}=
{{\bf E}a+{\bf B}b\over 2\sqrt{F^2+G^2}},\quad
{\partial a \over \partial {\bf B}}=-{\partial b \over \partial {\bf E}}=
{{\bf E}b-{\bf B}a\over 2\sqrt{F^2+G^2}}.
\end{equation}
substitution of which to~(\ref{PM_general}) finally leads to constitutive relations for
nonberturbative one-loop QED:
\begin{equation}\label{PM_explicit}
{\bf P}={\alpha \over 8\pi^2 (a^2+b^2)}\Big[I_1 {\bf E}+I_2{\bf B}\Big],
\quad {\bf M}=-{\alpha \over 8\pi^2 (a^2+b^2)}\Big[I_1{\bf B}-I_2 {\bf E}\Big],
\end{equation}
where for brevity we have used the notations for the integrals:
\begin{equation}\label{I1}
I_1=\int\limits_0^\infty \Big\{{ab\big[a \sinh(2sb)-b\sin(2sa)\big]\over 2 \sinh^2(sb)\sin^2(sa)}
-{2(a^2+b^2)\over 3 s}\Big\}e^{-s}ds,
\end{equation}
\begin{equation}\label{I2}
I_2=\int\limits_0^\infty \Big\{{ab\big[a \sin(2sa)+b \sinh(2sb)\big]\over 2 \sinh^2(sb)\sin^2(sa)}
-{(a^2+b^2)\over  s}\cot(sa)\coth(sb)\Big\}e^{-s}ds.
\end{equation}

It is easy to verify the correspondence to perturbative regime
which take place for relatively
weak fields $|{\bf E}|, |{\bf B}|\ll B_c$.
In this case, QED correction to Lagrangian
(\ref{Lagrange_main}) and the constitutive
relations (\ref{PM_explicit}) can be expanded in a
series by the small parameters
$a,b \ll 1$. Such expansion leads to:
\begin{equation}\label{L_decomp}
L_1={\alpha \over 8\pi B_c^2}
\Big[{\eta_1 ({\bf E}^2-{\bf B}^2)^2+2\eta_2 ({\bf E} {\bf B})^2}\Big],
\end{equation}
\begin{equation}\label{PM_decomp}
{\bf P}={\xi\over 2\pi}\Big\{\eta_1({\bf E}^2-{\bf B}^2)
{\bf E}+2\eta_2({\bf E} {\bf B}){\bf B}\Big\},\quad
{\bf M}=-{\xi\over 2\pi}\Big\{\eta_1({\bf E}^2-{\bf B}^2)
{\bf B}-2\eta_2({\bf E} {\bf B}){\bf E}\Big\},
\end{equation}
where $\xi=1/B_c^2$, $\eta_1=\alpha/45\pi$, and
$\eta_2=7\alpha/180\pi$ --  so called  post-Maxwellian parameters.
As the polarization and magnetization in (\ref{PM_decomp})
are  cubic on external fields, the electromagnetic waves
and charged particles propagation in perturbative QED vacuum
will possess the properties peculiar to
crystal optics with  cubic nonlinearity.
This approach leads to the predictions for
vacuum birefringence and dichroism~\cite{a6,a17,a18,a19},
optical non-reciprocity~\cite{a20},
light-ray bending~\cite{a23,a24} and Cherenkov-radiation~\cite{a21,a22}
in vacuum at presence of external electromagnetic field.
The expansion of these predictions
on nonperturbative QED regime  gives a new insight in understanding of
vacuum nonlinear electrodynamics and helps
to enforce the expectations in experimental manifestations.
In this paper we will focus only on vacuum birefringence expansion.
In order to do this, we will derive the dispersion relations for
electromagnetic wave propagating on background of strong
magnetic field.

\section{Electromagnetic wave propagation
in strong magnetic field}\label{Sec3}

\begin{figure}[tbp]
\centering
\includegraphics[width=.7\textwidth, clip]{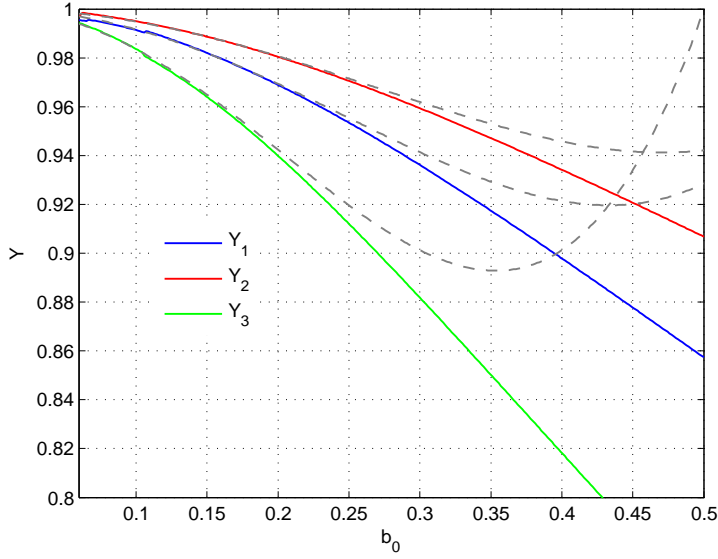}
\caption{\label{fig:1} An exact and  an approximate $Y$-functions comparison.}
\end{figure}

Let us  consider a weak electromagnetic wave ${\bf e}_w$,
${\bf b}_w$ propagating in strong
external magnetic field ${\bf B}_0$.
We suppose that the field intensity in the wave is
sufficiently weak, so $|{\bf e}_w|, |{\bf b}_w|\ll B_c, |{\bf B}_0|$.
As the vacuum nonlinear electrodynamics keeps the superposition principle,
the total field intensities  ${\bf B}={\bf B}_0+{\bf b}_w$ and ${\bf E}={\bf e}_w$
can be used in~(\ref{ab}) and in constitutive relations for
nonperturbative QED~(\ref{PM_explicit}).
Here we should take into account the weakness of the wave
and decompose these relations up to the leading order by
${\bf e}_w$ and ${\bf b}_w$. Such decomposition gives
linearized constitution relations:
\begin{eqnarray}\label{DH_wave}
{\bf D}&=&{\bf E}+4\pi{\bf P}= {\bf e}_w-
2\xi\Big\{\eta_1 Y_1 {\bf B}_0^2{\bf e}_w- 2\eta_2 Y_2 ({\bf B}_0
{\bf e}_w){\bf B}_0\Big\} \\
{\bf H}&=&{\bf B}-4\pi{\bf M}= {\bf B}_0+{\bf b}_w-2\xi\eta_1\Big\{Y_1 {\bf B}_0^2 {\bf b}_w+
2Y_3({\bf B}_0 {\bf b}_w){\bf B}_0+Y_1 {\bf B}_0^2 {\bf B}_0\Big\}, \nonumber
\end{eqnarray}
where we use linearized relations for the parameters $a$ and $b$
\begin{equation}\label{ab_line}
a={({\bf B}_0 {\bf e}_w)\over B_c B_0},\quad
b={B_0\over B_c}\Big\{1+{({\bf B}_0 {\bf b}_w)\over B_0^2}\Big\},
\end{equation}
introduce the notations for the integrals:
\begin{eqnarray}\label{Y_exact}
 Y_1  &=& -{45\over 4b_0^2}
\int\limits_0^\infty {e^{-z/ b_0}\over z^2}
\Big\{\coth(z)-{z\over \sinh^2(z)}-{2z\over 3}\Big\}dz, \\
 Y_2  &=& {45\over 14 b_0^2}
\int\limits_0^\infty {e^{-z/ b_0}\over z^2}
\Big\{{2z^2-3\over 3}\coth(z)+{z\over \sinh^2(z)}\Big\}dz, \\
  Y_3 &=& -{45\over 8 b_0^2}
\int\limits_0^\infty {e^{-z/ b_0}\over z^2}
\Big\{{2z^2 \coth(z)-z\over \sinh^2(z)}-\coth(z)\Big\}dz,
\end{eqnarray}
and the dimensionless parameter ${\bf b}_0={\bf B}_0/ B_c$.
The correspondence to the perturbative regime can be
obtained when $B_0\ll B_c$. This leads to the
asymptotic expansion of the integrals:
\begin{equation}\label{Int_expans}
Y_1=1-{6\over 7}b_0^2+{16\over 7}b_0^4+\cdots ,\quad
Y_2=1-{26\over 49}b_0^2+{176\over 147}b_0^4+\cdots ,\quad
Y_3=1-{12\over 7}b_0^2+{48\over 7}b_0^4+\cdots.
\end{equation}
The graphs for an exact~(\ref{Y_exact})
and an approximate~(\ref{Int_expans}) $Y$
dependence on magnetic field strength
are represented on Figure~\ref{fig:1}.
Approximate functions are
marked on the graph by the gray line.
As it can be seen, even in low field values $b_0\ll1$ an approximate description
can cause significant inaccuracies
whose rectification will require new terms  in the expansion~(\ref{Int_expans}).

Also it should be noted that, as the unity is the
leading term in all of the listed above
expansions, the linearized constitutive relations~(\ref{DH_wave})
contain perturbative relations~(\ref{PM_decomp})
in low field limit. Besides, the 	
transition from the perturbative QED to
nonperturbative regime is in
replacement of "post-Maxwellian" constants $\eta_1$
and $\eta_2$ on the functions $\zeta_1=\eta_1Y_1(b_0)$,
$\zeta_2=\eta_2Y_2(b_0)$,
$\zeta_3=\eta_1Y_3(b_0)$ which depend on
the external field strength.
To obtain dispersion relations we use the eikonal
approximation and represent
the wave field in form:
${\bf e}_w={\bf e}\exp{\{i S({\bf r},t)\}},$
${\bf b}_w={\bf b}\exp{\{i S({\bf r},t)\}},$
where ${\bf e}$ and ${\bf b}$ are the amplitudes
and $S$ is the eikonal.
As usual for the eikonal
approximation, we will suppose that amplitude variations along the
wave propagation ray are small and can be neglected in calculations.
Substitution of ${\bf e}_w$, ${\bf b}_w$ and  constitutive relations
(\ref{DH_wave}) to the ordinary equations of continuous media electrodynamics
leads to homogeneous equations on wave field components:
\begin{equation}\label{Eic_eq}
\Pi_{\alpha \beta}{e}_w^\beta=0,
\end{equation}
where indexes enumerate the cartesian components of the
electric field vector $\alpha,\beta=1\ldots3$ and
$\Pi_{\alpha \beta}$ is the polarization tensor:
\begin{eqnarray}\label{Polariz_ten}
\Pi_{\alpha\beta}=\Big\{\Big(({\bf \nabla } S)^2-
({\partial_0 S})^2\Big)\delta_{\alpha\beta}-
{\partial_\alpha S}{\partial_\beta S}\Big\}\times
\Big\{1-2\zeta_1 b_0^2\Big\} \\
-4\zeta_3\Big[{\bf \nabla }S {\bf b}_0\Big]_{\alpha}
\Big[{\bf \nabla }S {\bf b}_0\Big]_{\beta}
-4\zeta_2({\partial_0 S})^2({\bf b}_0)_{\alpha}({\bf b}_0)_{\beta}, \nonumber
\end{eqnarray}
in which the following notations were used:
$\partial_0=\partial /\partial (ct)$ and $\partial_\alpha$ denotes
spatial coordinate $x^{\alpha}$ derivative, $\delta_{\alpha\beta}$ --
is the Kronecker symbol, $\nabla$ -- is the gradient
operator, ${\bf b}_0={\bf B}_0/B_c$ and the square
brackets refer to the vector cross-product.
In should be noted that our result  for polarization tensor is close to the similar one obtained in~\cite{new1}.
The existence of nontrivial solutions of (\ref{Eic_eq}) requires
$\mbox{det}\|\Pi_{\alpha\beta}\|=0$ which finally leads to
dispersion relations for the electromagnetic
wave at the eikonal approximation:
\begin{eqnarray}\label{Dispers}\nonumber
\Big\{({\bf \nabla } S)^2-({\partial_0 S})^2+4\zeta_3\Big[({\bf b}_0 {\bf\nabla}S)^2
&-&{ b}_0^2({\bf \nabla}S)^2\Big]+2\zeta_1{ b}_0^2\Big[
({\partial_0 S})^2-({\bf\nabla}S)^2\Big]\Big\} \\  \nonumber
\times \Big\{({\bf\nabla}S)^2-
({\partial_0 S})^2+4\zeta_2\Big[({\bf b}_0 {\bf \nabla}S)^2
&-&{ b}_0^2({\partial_0 S})^2\Big]+2\zeta_1{ b}_0^2\Big[
({\partial_0 S})^2-({\bf\nabla}S)^2
\Big]\Big\}  \\
&\times&\Big\{2\zeta_1{ b}_0^2-1\Big\}=0.
\end{eqnarray}

Multiplicative structure of obtained dispersion
relations points to vacuum birefringence retention
even in case of one-loop nonperturbative QED.
The first factor in (\ref{Dispers}) corresponds to
the dispersion law for a normal wave
polarized perpendicular to external magnetic field ($\bot$-mode),
whereas the second multiplier describes
the mode polarized along the field ${\bf B}_0$ ($||$-mode).
The last factor in (\ref{Dispers}) does not depend on the
wave parameters and its equality to zero is expected at
huge magnetic field intensities $B_0\sim B_c \exp{(1/\alpha)}$.
Just for these values QED vacuum instability induced by  magnetic
field was predicted~\cite{a25}.
To avoid such a regime in future, we will
consider field intensities much closer to $B_c$.

Now let us investigate the properties of normal waves in more detail,
and use the obtained dispersion relations
in strengthening some expectations for vacuum birefringence detection
in experiment.

\section{QED vacuum birefringence extension on nonperturbative regime}\label{Sec4}

There are two traditional approaches to the interpretation
of the dispersion relations
for vacuum nonlinear electrodynamics.
The first one comes from the representation of the vacuum
as a continuous media. The vacuum birefringence in this case
is explained by the normal modes
refraction indexes mismatch $n_{\bot}\neq n_{||}$.
The second approach assumes that the electromagnetic
wave propagates in the space-time with the
effective geometry following from the dispersion relations.
This interpretation explains the vacuum birefringence
due to the difference between the effective space-time metric
tensors $G^{ik}_{\bot}\neq G^{ik}_{||}$ for each normal mode.
For completeness, we use each of these
approaches in the analysis of obtained dispersion relations.

For description in terms of refractive indexes
one should substitute eikonal $S(t,{\bf r})=\omega t-({\bf k}{\bf r})$
to dispersion relations (\ref{Dispers}),
and  take into account the relation between the wave vector
${\bf k}$ and frequency $\omega$ which is ordinary for homogeneous
wave in continuous media: ${\bf k}=\omega n {\bf q}/c $,
where  $n$ is the refraction index and ${\bf q}$ is the
unity vector in wave propagation direction.
Such substitution gives  explicit
expressions for the normal modes refraction indexes:
\begin{equation}\label{N_exact}
n_{\bot}^2=1+{4\zeta_3 b_0^2\sin^2{\theta}
\over 1-2b_0^2[\zeta_1+2\zeta_3\sin^2{\theta}]}, \qquad
n_{||}^2=1+{4\zeta_2b_0^2\sin^2{\theta}
\over 1-2b_0^2[\zeta_1-2\zeta_2\cos^2{\theta}]},
\end{equation}
where $\theta$ is an angle between the wave vector ${\bf k}$ and
the magnetic field ${\bf B}_0$.
These expression refine the results obtained earlier~\cite{a26}
in which the dependance on angle $\theta$ is more simple
and the terms in denominator are neglected:
\begin{equation}\label{N_apr}
n_{\bot}^2\approx1+{4\zeta_3  b_0^2\sin^2{\theta}}, \qquad
n_{||}^2\approx1+{4\zeta_2 b_0^2\sin^2{\theta}},
\end{equation}
which are close to results of~\cite{a16, a26}
obtained on the basis of a purely quantum
approach with fixed selection of the gauge.
As the angle dependance in the exact expressions~(\ref{N_exact})
is different, it becomes possible to figure out are there any
conditions under which $n_{\bot}=n_{||}$ and the birefringence is suppressed.
The equality of the refractive indexes leads to
the relation which is valid for any angle $\theta$:
\begin{equation}\label{Birefr_suppr}
(\zeta_3-\zeta_2)(1-2\zeta_1b_0^2)+4\zeta_3\zeta_2b_0^2=0.
\end{equation}
However, the numerical calculations show that
there are no solutions for this equation at
any arbitrary magnetic field close to $B_c$
so the vacuum birefringence remains.

\begin{figure}[tbp]
\includegraphics[width=.49\textwidth,clip]{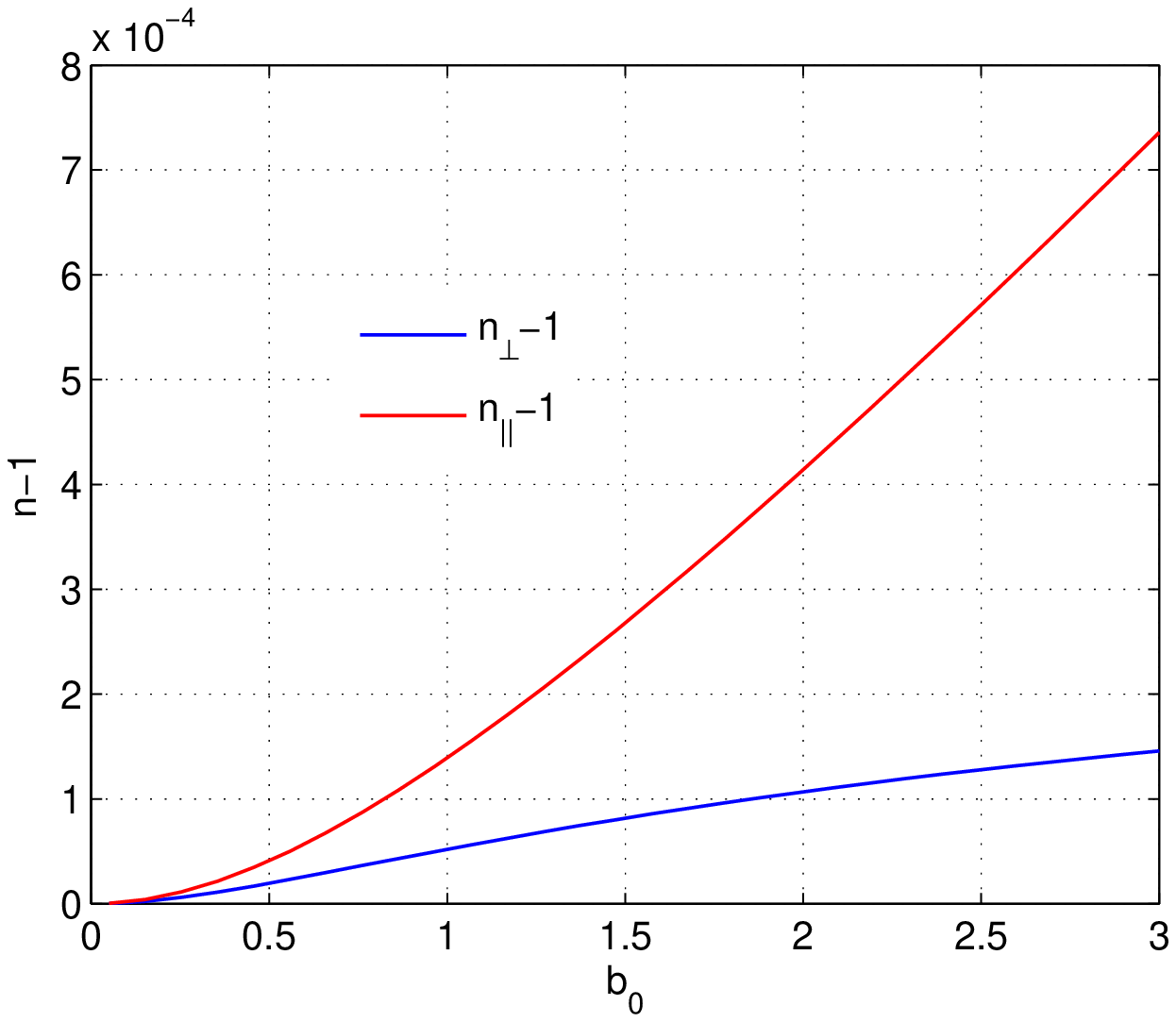}
\hfill
\includegraphics[width=.49\textwidth]{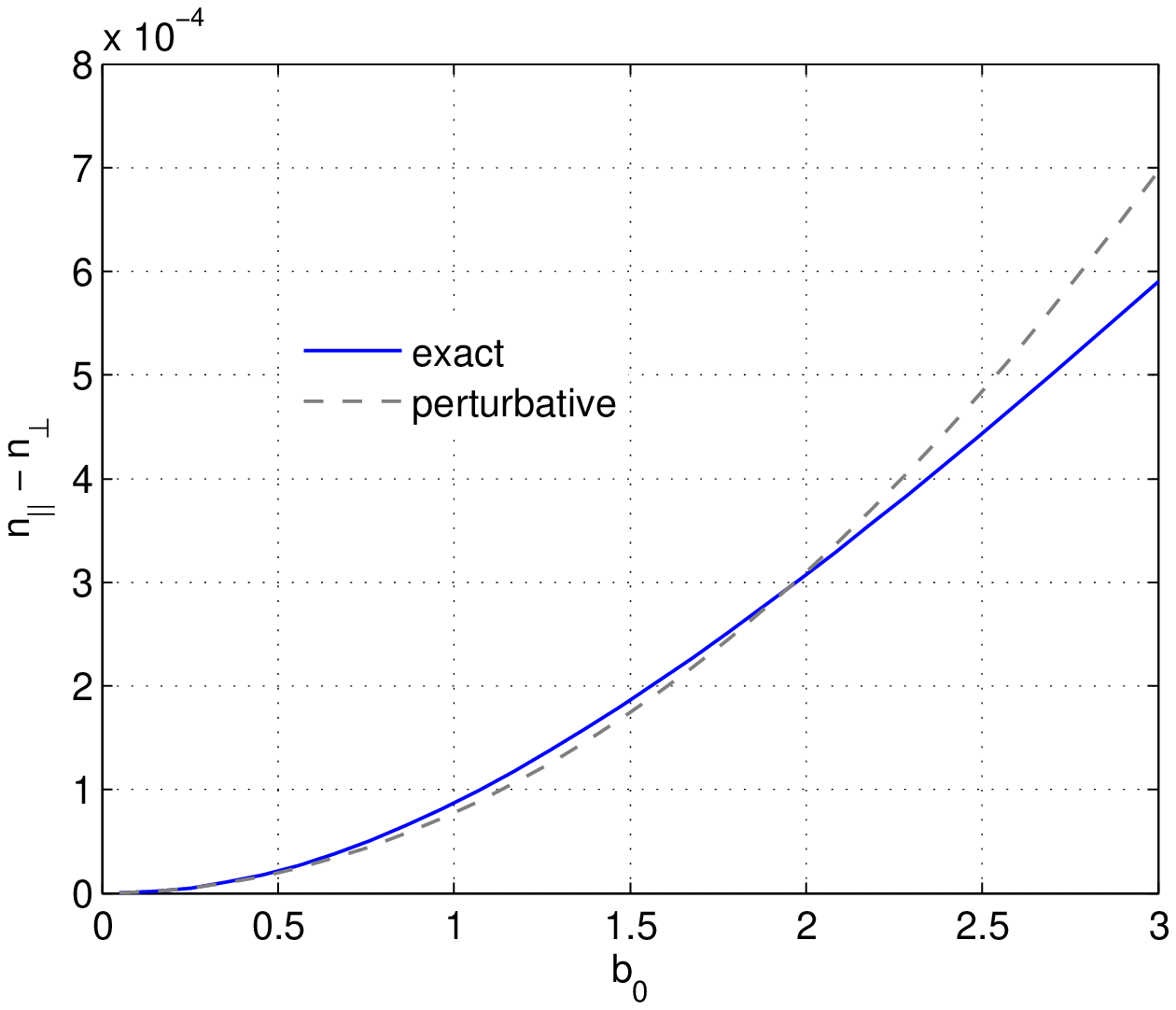}
\caption{\label{fig:2} The refractive indexes and their difference
depending on the magnetic field value $b_0=B_0/B_c$.}
\end{figure}

To analyze the dependence of  refractive indexes
on the magnetic field value, we will choose $\theta=\pi/2$,
which is more valuable for experimental research
because in this case  indexes are maximal.
As it follows from  numerical calculations,
the refractive index $n_{||}$ increases almost linearly
in $1<b_0<100$ and shows nonlinear growth
in wider field ranges.
Whereas the $n_{\bot}$ tends to saturation
at the value  $(n_{\bot})_{sat}-1\approx 4\cdot 10^{-4}$ and
ceases to depend on the field strength.
The dependance of $n_{\bot,||}-1$ for $b_0<3$ is
represented on the left graph in Figure~\ref{fig:2}.
The right graph shows the exact difference $n_{||}-n_{\bot}$
coming from~(\ref{N_exact}) and the same difference
following from the perturbative QED:
$n_{||}-n_{\bot}\approx2(\eta_2-\eta_1)b_0^2$,
which is marked on the graph with a gray line.
Despite the proximity of perturbative description,
there is a good accordance to the exact result
up to the field values $b_0\approx2$.
This indicates that more then order
discrepancy between perturbative QED birefringence prediction
and the experimental result obtained in PVLAS~\cite{a8}
is not associated with the inaccuracy of perturbative
description and should have a more
profound physical reason.

Another traditional approach to
the dispersion relations interpretation
is based on effective geometry representation.
It assumes that the wave propagates in
curved space-time, which geometry
depends on external magnetic field.
The dispersion relations (\ref{Dispers}), now are interpreted as
Hamilton-Jacobi equation for a massless
particle in the effective space-time
with  the metric tensor $G^{ik}_{\bot,||}$ correspondent to each
normal mode:
\begin{equation}\label{HJ_equation}
\Big[G^{ik}_{\bot}{\partial S \over \partial x^i}{\partial S \over \partial x^k}\Big]\times
\Big[G^{mn}_{||}{\partial S \over \partial x^m}{\partial S \over \partial x^n}\Big]=0.
\end{equation}
As it follows from (\ref{Dispers}), the components of the
effective metric tensor  take the form:
\begin{eqnarray}\label{Eff_MT}
G^{00}_{\bot}&=&g^{00},
\quad G^{\alpha\beta}_{\bot}=g^{\alpha\beta}+
{4\zeta_3\over 1-2\zeta_1 b_0^2} \times (b_0^2\delta^{\alpha\beta}-b_0^\alpha b_0^\beta), \\
G^{00}_{||}&=&g^{00},
\quad G^{\alpha\beta}_{||}=g^{\alpha\beta}+
{4\zeta_2\over
1-2\zeta_1 b_0^2+4\zeta_2  b_0^2}\times(b_0^2\delta^{\alpha\beta}-b_0^\alpha b_0^\beta),
\end{eqnarray}
where $g^{ik}$ is Minkowski space-time metric tensor
and the Greek indexes enumerate the spatial
coordinates, so $\alpha, \beta=1..3$. It is easy to verify that
in low-field limit $b_0\ll1$ the expressions (\ref{Eff_MT})
take the form of perturbative QED effective metric tensor
obtained in~\cite{a11}:
\begin{equation}\label{Eff_pert}
G^{ik}_{\bot,||}\approx g^{ik}+4\eta_{1,2}
(b_0^2\delta^{\alpha\beta}-b_0^\alpha b_0^\beta).
\end{equation}
Such correspondence allows us to extend some
predictions for vacuum birefringence manifestations
obtained on base of  perturbative
metric tensor (\ref{Eff_pert}) to non-perturbative
region by simple replacement of parameters:
\begin{equation}\label{PM_rep}
\eta_1\to {\zeta_3\over 1-2\zeta_1 b_0^2},
\quad \eta_2 \to {\zeta_2 \over 1-2\zeta_1 b_0^2+4\zeta_2 b_0^2}.
\end{equation}
For instance, such extension is especially actual
for the pulsars and magnetars which are
one the most attractive objects for QED regime tests.
The magnetic fields of such astrophysical
sources can significantly exceed critical limit $b_0=B_0/B_c\gg1$.
One of observable manifestations of the vacuum birefringence
in pulsar neighbourhood is related to normal mode delay
for hard X-ray and gamma- radiation pulses passing near the pulsar.
Because of the vacuum birefringence, the propagation velocity
of ${\bot}$-mode is greater then ${||}$-mode, so it will
reach to the detector earlier. So the leading part of the pulse
coming from the X-ray source to the detector will be
linearly polarized due to the ${\bot}$-mode.
This part of the pulse will have a time duration $\Delta t$.
After this time the ${||}$-mode will
reach the detector and the pulse polarization state
will change to elliptical. The maximal estimation for
the delay $\Delta t$ in perturbative QED
was obtained in~\cite{a11}:
\begin{equation}\label{DelT_pert}
\Delta t={123\pi(\eta_2-\eta_1)b_0^2R_s\over 128c},
\end{equation}
where $R_s$ is the pulsar radius, and $c$ is the speed of light
in Maxwell vacuum. Birefringence expansion (\ref{PM_rep})
allows us to estimate the order of magnitude
for the time delay $\Delta t$ in nonperturbative regime:
\begin{equation}\label{DelT_Npert}
\Delta t\simeq {b_0^2 R_s \over c}\Big[{\zeta_3\over 1-2\zeta_1 b_0^2}-
{\zeta_2 \over 1-2\zeta_1 b_0^2+4\zeta_2 b_0^2}\Big].
\end{equation}

\begin{figure}[tbp]
\centering
\includegraphics[width=.7\textwidth, clip]{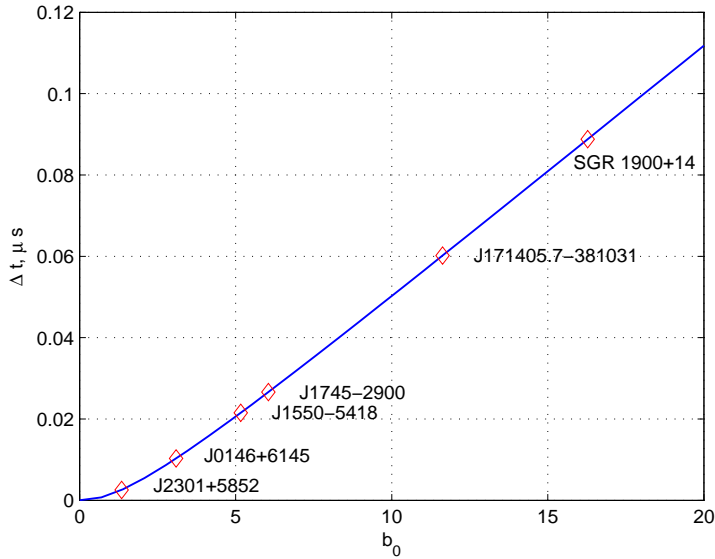}
\caption{\label{fig:3} Time delay between ${\bot}$ and ${||}$-modes arrival to detector.}
\end{figure}

The dependence of the time delay on magnetic field strength
and its estimates for some pulsars and magnetars are represented on
Figure~\ref{fig:3}. For the estimates we use the pulsar data from
McGill~\cite{a27} and ATNF~\cite{a28} catalogs
and also suppose the pulsar radius equal to $R_s=10 km$.
The delay value varies in hundredth of microseconds and
this is sufficient for contemporary timing measurements.
Furthermore the estimate can be enforced for the
unique object J1808-2024 with the field strength $B_0\sim 2.06\cdot 10^{15} G$
for which the delay can reach $\Delta t\sim 0.3\mu s$.
Also it should be noted that there is almost
linear dependence between delay and field strength,
instead of quadratic, typical for delay in
perturbative regime (\ref{DelT_pert}), therefore
the direct interpolation of perturbative description
on the region $b_0>1$  will lead to
overestimation for the delay.

\section{Conclusion}\label{Sec5}
In this paper we have investigated  expansion
of vacuum birefringence on nonperturbative regime of QED.
The obtained constitutive relations~(\ref{DH_wave})
for  weak electromagnetic
wave propagating on background of a strong magnetic field
indicate that this expansion consists of replacing of perturbative constants
$\eta_1,\eta_2$ by three functions $\zeta_1, \zeta_2, \zeta_3$,
which  depend on the  external field strength.
For  vacuum birefringence description we have used a semiclassical
approach in terms of the wave field strength
which gives gauge independent results unlike
most of the other calculations performed in the fixed gauge selection.
In such approach  the polarization
tensor (\ref{Polariz_ten}) and dispersion
relations (\ref{Dispers}) were obtained and both interpreted
in terms of refractive indexes and the
effective space-time geometry. The refractive indexes
interpretation confirmed the results obtained earlier
by other authors on the base of quantum field theory methods.
The comparison between the perturbative and nonperturbative
refractive indexes, as expected,
reveals an insignificant difference at weak field values $b_0\ll1$,
and this indicates that inaccuracy in QED theoretical
description can not be a cause of discrepancy
detected in the PVLAS experiment~\cite{a9}.
Another feature observed in the
refractive indexes analysis is the possible saturation of $n_{||}$ at
the strong external field values $b_0\gg1$.
Unfortunately, this feature can not
be verified in conditions of the terrestrial
facilities, however the astrophysical sources
such as pulsars and magnetars provide a
wider opportunities in QED features investigations.
It is more convenient to use
an effective geometry formalism for vacuum birefringence
description in the neighbourhood of  such field sources.
The dispersion relations interpretation in terms of the
effective geometry~(\ref{Eff_MT}) allowed us to extend
the predictions for the normal mode relative delay
in the hard radiation propagating near the pulsar.
The estimates for  such delay for a different pulsars, show
its detectability with  contemporary experimental
technique. Moreover, it was found out that
the dependence between the delay and the field
value predicted by nonperturbative QED is close to linear (Figure \ref{fig:3})
and can be roughly expressed as $\Delta t=(5.8b_0-7)\times 10^{-3}\mu{s}$
for $2<b_0<100$.
This dependence differs from the quadratic one following from
the direct perturbative QED prediction extrapolation to the strong field region.
As the perturbative QED is not valid for such  field values,
it gives overestimated result for delay. This feature can be noted
in planning of  future astrophysical missions aimed at
the QED effects investigation in the pulsar fields,
such as XIPE~\cite{a30}.

\end{document}